# Two-dimensional model of interaction of a non-relativistic particle with scalar mesons in the strong coupling limit

A. Shurgaia

I. Javakhishvili State University, A. Razmadze Mathematical Institute

1 M. Alexidze str. 0171 Tbilisi, Georgia

Bogolyubov transformations are introduced into the nonrelativistic model of particle interaction with scalar mesons. Within the framework of the generalized Hamiltonian formalism developed by Dirac, a translation-invariant perturbation theory in inverse powers of the coupling constant is constructed. Transmission and reflection coefficients are obtained, the S-matrix is written in a holomorphic representation, and the matrix element is calculated.

## Introduction

Collective coordinates method in strong coupling theory [1] allows one to construct a consistent scheme of successive approximations for the energy of the wave function of a system, and the invariance properties can be described exactly. The development and generalization of this method for systems invariant with respect to an arbitrary group of transformations are given in [2, 3]. The application of the method to various models is the subject of [4].

This method (also called the method of collective variables) has proven to be effective in nonlinear models of field theory that admit particle-like solutions of classical equations. The canonical Hamiltonian approach is given in [5], and in [6] it is developed for path integrals.

Bogolyubov transformations introduce additional variables into the theory. As a result, we arrive at singular Lagrangian. The Hamiltonian formalism for such Lagrangian was constructed by Dirac [7]. Based on this formalism, the method of collective variables was developed in [8, 9] for Lagrangian with arbitrary symmetry. In this paper, based on [8], the simplest translation-invariant model of the interaction of a nonrelativistic particle with scalar mesons in two-dimensional space-time is considered. A similar problem in the canonical formalism was investigated in [10]. The Lagrangian of the system has the form

$$L = \frac{M\dot{X}^2}{2} + \frac{1}{2}\int dx [\dot{\Phi}^2(x,t) + \Phi(x,t)(\Delta - \mu^2)\Phi(x,t)] -$$
$$-g \int dx K(x - X)\,\Phi(x,t), \tag{1}$$

where $X$ and $\Phi(x,t)$ are the canonical coordinates of the particle and the field. $M$ is the mass of the particle, and since we are interested in the case where the coupling constant is large, we will assume that $M = g^2 M_0$, $K(x)$ is an even cutoff function that decreases rather rapidly as $|x| \to \infty$.

The Hamiltonian of the system has the form

$$H = \frac{p_x^2}{2} + \frac{1}{2}\int dx [\Pi^2(x,t) + \Phi(x,t)(-\Delta + \mu^2)\Phi(x,t)] -$$
$$+g \int dx K(x - X)\,\Phi(x,t), \tag{2}$$

## 1. Transformation of variables

The classical equation

$$(-\partial_t^2 + \partial_x^2 + \mu^2)\Phi(x,t) = -gK(x)\Phi(x,t)$$

for the field function $\Phi(x,t)$ we will consider for a fixed coordinate of the particle. We will assume that the source function is translation-invariant. Then the requirement of invariance of the equation with respect to two-dimensional shifts allows us to seek solutions in the form $\Phi(x,t) \to \Phi(x - a(t))$ where $a(t)$ is a linear function of time. Such functions satisfy the equation

$$(v^2 - 1)\ \Phi''(x,t) + \mu^2 \Phi(x,t) = -gK(x)\Phi(x,t). \tag{3}$$

Here $v = \dot{a}(t)$ and the shift $x \to x - a(t)$ is made. It is clear that the solution is proportional to the coupling constant $g$. $\Phi_c(x) = g\varphi_0(x)$, where

$$\varphi_0(x) = -\frac{1}{2\mu\sqrt{1-v^2}} \int dy K(y) \exp\left(-\frac{\mu|x-y|}{\sqrt{1-v^2}}\right). \tag{4}$$

The solution of the equation of motion for the coordinate of a particle $X$ is proportional to the zeroth power of $g$. In view of all this, we introduce the following transformations of the coordinate of the particle $X$ and the field $\Phi(x,t)$:

$$X = a + \frac{\lambda}{g}, \quad \Phi(x,t) = \overline{\Phi}(x - a(t), t). \tag{5}$$

Thus, the canonical variables of the system become $a, \lambda$ and $\Phi(x - a(t), t)$. $\lambda$ will describe the quantum effect of the interaction of the particle with the field, and the quantity $a$ will describe the motion of the system as a whole. Lagrangian (1) after transformation (5) takes the form:

$$L = \frac{M}{2}\left(\dot{a} + \frac{\dot\lambda}{g}\right)^2 + \frac{1}{2}\int dx\left[\left(\dot{\overline\Phi}(x.t) - \dot{a}\overline\Phi'(x,t)\right)^2 + \right.$$

$$\left. + \overline\Phi(x,t)(\partial_x^2 - \mu^2)\overline\Phi(x,t)\right] - \int dx K\left(x - \frac{\lambda}{g}\right)\overline\Phi(x,t). \tag{6}$$

The canonical momenta

$$p_\lambda = \frac{\partial L}{\partial \dot\lambda} = \frac{M}{g}\left(\dot{a} + \frac{\dot\lambda}{g}\right),$$

$$p_a = \frac{\partial L}{\partial \dot a} = M\left(\dot a + \frac{\dot\lambda}{g}\right) - \left[\dot{\overline\Phi}(x.t) - \dot{a}\overline\Phi'(x,t)\right]\overline\Phi'(x,t),$$

$$\overline{\Pi}(x,t) = \frac{\delta L}{\delta \dot{\overline{\Phi}}(x,t)} = \dot{\overline{\Phi}}(x,t) - \overline{\Phi}'(x,t)$$

obey the following constraint:

$$\psi(p_a, p_\lambda, \overline{\Pi}(x,t), \overline{\Phi}(x,t)) = p_a - g p_\lambda + \int dx\, \overline{\Pi}(x,t)\overline{\Phi}'(x,t) = 0. \qquad (6)$$

Taking into account (6), for the Hamiltonian (2) we obtain

$$H = \frac{g^2 p_\lambda^2}{2M} + \frac{1}{2}\int dx [\overline{\Pi}^2(x,t) + \overline{\Phi}(x,t)(-\partial_x^2 + \mu^2)\overline{\Phi}(x,t) +$$

$$+ g \int dx\, K\left(x - \frac{\lambda}{g}\right) \overline{\Phi}(x,t). \qquad (7)$$

Let us introduce the transformation:

$$\overline{\Phi}(x,t) = g\varphi_0(x) + \varphi(x,t), \quad \overline{\Pi}(x,t) = g\pi_0(x) + \pi(x,t). \qquad (8)$$

where $\pi_0(x) = -v\varphi_0'$. The constraints (6) now can be rewritten as

$$F + \int dx\, \pi(x,t)\varphi_0'(x) = 0. \qquad (9)$$

$$F = p_a - gp_\lambda + g^2 \int dx\, \pi_0(x)\varphi_0'(x) + g\int dx\, \pi_0(x)\varphi'(x,t) +$$

$$\int dx\, \pi(x,t)\varphi'(x,t).$$

The explicit form of the connection allows us to break the momentum $\pi(x,t)$ into transverse $\pi_t(x,t)$ and longitudinal $\pi_l(x,t)$ relative to $\varphi_0'(x)$ components $\pi(x,t) = \pi_t(x,t) + \pi_l(x,t)$: $\int dx\, \varphi_0'(x)\pi_t(x,t) = 0$. Let's define $\pi_l(x,t)$ as follows:

$$\pi_l(x,t) = -\frac{\varphi_0'(x)}{gn_0} F \quad \text{with} \quad n_0 = \int dx\, \varphi_0'(x)^2. \qquad (10)$$

Substituting (10) into (9), we find an explicit expression for $F$. Therefore, one obtains for the momentum $\pi(x,t)$

$$\pi(x,t) = \pi_t(x,t) - \frac{\varphi_0'(x)}{gn_0}\left(1 + g\frac{n}{n_0}\right)^{-1}(p_a - gp_\lambda) +$$

$$g^2 \int dx\, \pi_0(x)\varphi_0'(x) + g\int dx\, \pi_0(x)\varphi'(x,t) +$$

$$\int dx\, \pi_t(x,t)\varphi'(x,t). \qquad (11)$$

with $n = \int dx\, \varphi'(x,t)\varphi_0'(x)$.

Constraint now becomes

$$\int dx \pi_t(x,t)\varphi_0'(x) = 0. \qquad (12)$$

Substituting (11) into (7) taking into account (8), we obtain for the Hamiltonian:

$$H = \frac{g^2 p_\lambda^2}{2M} + \frac{1}{2}\int dx\, \pi_t^2(x,t) + \frac{g^2}{2}\int dx\, \pi_0^2(x) +$$

$$+ g\int dx\, \pi_0(x)\pi_t(x,t) + \frac{1}{2g^2 n_0}\left(1 + \frac{n}{gn_0}\right)^{-2}\left(p_a - gp_\lambda + \right.$$

$$+ g^2\int dx\, \pi_0(x)\varphi_0'(x) + g\int dx\, \pi_0(x)\varphi'(x,t) + \int dx\, \pi_t(x,t)\varphi'(x,t))^2 -$$

$$- \frac{1}{n_0}\left(1 + \frac{n}{gn_0}\right)^{-1}\left(\int dx\, \pi_0(x)\varphi'(x,t)\right)\left(p_a - gp_\lambda + g^2\int dx\, \pi_0(x)\times\right.$$

$$\times\varphi_0'(x) + g\int dx\, \pi_0(x)\varphi'(x,t) + \int dx\, \pi_t(x,t)\varphi'(x,t) +$$

$$+ \frac{1}{2}\int dx\, (g\varphi_0(x) + \varphi(x,t))(-\Delta + \mu^2)(g\varphi_0(x) + \varphi(x,t)) +$$

$$+ g\int dx\, K\left(x - \frac{\lambda}{g}\right)(g\varphi_0(x) + \varphi(x,t)).$$

(13)

However, the Hamiltonian (7) is not uniquely defined from the very beginning. To define it uniquely, the following additional constraint has to be introduced:

$$\chi = \int dx\, \varphi_0'(x)\varphi(x,t) = 0. \qquad (14)$$

Now the action of the system (taking into account constraints) is

$$I = \int_{t_i}^{t_f} dt\, (p_a \dot{a} + p_\lambda \dot{\lambda} + \int dx \pi_t(x,t)\dot{\varphi}(x,t) - H.$$

Next, using the generalized Hamiltonian formalism and Pousson bracket $\{\psi, \chi\} = -n_0$, we find the Dirac brackets for the dynamic variables:

$$\{a, p_a\}_D = \{\lambda, p_\lambda\}_D = 1, \qquad (15)$$

$$\{\varphi(x,t), \pi_t(y,t)\}_D = \delta(x-y) - \frac{1}{n_0}\varphi_0'(x)\varphi_0'(xy) \qquad (16)$$

All other brackets are zero.

The Hamiltonian of system (13) depends on $a$ only in combination with $p_a$. The equation of motion for this quantity leads explicitly to the conservation of $p_a$, in particular $\{p_a, H\}_D = 0$. On the other hand, writing the total momentum of the system in new variables, we find the quantity $p_a$ (in new dynamic variables) to be the total momentum of the system:

$$p_a = P = gp_\lambda - \int dx \bar{\Pi}(x,t)\bar{\Phi}'(x,t). \tag{17}$$

## 2. Perturbation theory

In the Hamiltonian (13) we make the substitution $p_a \to gp_{a0}$, $p_\lambda \to p_\lambda + gC$ and expand it into a series in powers of g: $H = g^2 H_0 + gH_1 + H_0 + \cdots$, in which

$$H_2 = \frac{v^2 n_0}{2} + \frac{1}{2}\int dx \varphi_0(x)[(v^2-1)\varphi_0''(x) + \mu^2 \varphi_0(x)] +$$
$$\int dx K(x)\varphi_0(x) + \frac{(p_{a0}-C)^2}{2n_0} + \frac{C^2}{2M_0},$$

$$H_1 = \int dx \varphi(x,t)\left[\left(\frac{(p_{a0}-C)^2}{n_0^2} - 1\right)\varphi_0''(x) + \mu^2 \varphi_0(x) + K(x)\right] +$$
$$+ \frac{p_\lambda C}{M_0} - \frac{p_\lambda(p_a - C)}{n_0}.$$

$$H_0 = \frac{p_\lambda^2}{2m} + \frac{m\omega^2 \lambda^2}{2} + \int dx[\pi_t^2(x,t) + \varphi(x,t)(-\partial_x^2 + \mu^2)\varphi(x,t) -$$
$$-\lambda \int dx K'(x)\varphi(x,t) - \frac{p_{a0}-C}{n_0}\int dx\, \pi_t(x,t)\varphi'(x,t) +$$
$$\frac{2p_\lambda(p_{a0}-C)}{n_0^2}\int dx + \frac{3(p_{a0}-C)^2}{2n_0^3}\left(\int dx \varphi_0'(x)\varphi'(x,t)\right)^2.$$

Here the following notations are introduced: $\frac{1}{m} \equiv \frac{1}{n_0} + \frac{1}{M_0}$, $m\omega^2 \equiv \int dx\, K''(x)\varphi_0(x)$. Taking into account (3), we select the number C as follows: $(p_{a0}-C)/n_0 \equiv C/M_0 \equiv v$. It follows that $C = vM_0$ and the total mometnum of the system $p_a = g^2(M_0 + n_0)$. All thes lead for $H_2$ to the following expression:

$$H_2 = \frac{v^2 n_0}{2} + \frac{v^2(M_0 + n_0)}{2}\frac{1}{2}\int dx K(x). \tag{18}$$

Now, $H_1$ is identically zero and in the system where $v = 0$ one obtains for $H_0$:

$$H_0 = \frac{p_\lambda^2}{2m} + \frac{m\omega^2}{2} + \int dx[\pi_t^2(x,t) + \varphi(x,t)(-\partial_x^2 + \mu^2)\varphi(x,t) - \\ -\lambda \int dx K'(x)\varphi(x,t) \quad (19)$$

Let's derive the equations of motion in terms of Dirac's brackets.

$$\dot{\lambda} = \{\lambda, H_0\} = p_\lambda/m, \quad (20)$$

$$\dot{p}_\lambda = \{p_\lambda, H_0\}_D = -m\omega^2 \lambda + \int dx K'(x)\varphi(x,t),$$

$$\dot{\varphi}(x,t) = \{\varphi(x,t), H_0\}_D = \pi_t(x),$$

$$\dot{\pi}(x,t) = \{\pi(x,t), H_0\}_D = -(-\partial_x^2 + \mu^2)\varphi(x,t) +$$

$$\lambda\left\{K'(x) + \frac{m\omega^2}{n_0}\varphi_0'(x)\right\} + \frac{\varphi_0'(x)}{n_0}\int dx \varphi_0'(x)(-\partial_x^2 + \mu^2)\varphi(x,t).$$

After exclusion of $p_\lambda$ and $\pi(x,t)$ we find:

$$\ddot{\lambda} = -\omega^2 \lambda + \frac{1}{m}\int dx\, K'(x)\varphi(x,t). \quad (21)$$

$$\ddot{\varphi}(x,t) = -(-\partial_x^2 + \mu^2)\varphi(x,t) + \lambda\left\{K'(x) + \frac{m\omega^2}{n_0}\varphi_0'(x)\right\} +$$

$$\frac{\varphi_0'(x)}{n_0}\int dx \varphi_0'(x)(-\partial_x^2 + \mu^2)\varphi(x,t).$$

We'll look for solution for $\lambda$ and $\varphi(x,t)$ as follows:

$$\lambda = \lambda_k \exp(-i\omega t), \quad \varphi(x,t) = \varphi_k(x)\exp(-i\omega t).$$

One obtains on substituting into (21):

$$m(\omega^2(k) - \omega^2)\lambda_k = U(k) \quad (22)$$

$$(-\partial_x^2 + k^2)\varphi_k(x) = \lambda_k\left(K'(x) + \frac{m\omega^2}{n_0}\varphi_0'(x)\right)\frac{\varphi_0'(x)}{n_0}U(k),$$

$$\omega^2(k) = k^2 + \mu^2, \quad U(k) = \int dx \varphi_0'(x)(-\partial_x^2 + \mu^2)\varphi_k(x).$$

The solutions of the system exist at any real value of $k$ $(-\infty < k < \infty)$ and are [10]:

$$\lambda_\kappa^\pm = -Nk\frac{\varphi_0(k)}{n_0\Delta^\pm(k)}, \quad (23)$$

$$\varphi_k^\pm(x) = N\left\{e^{ikx} + \frac{ik\varphi_0(k)}{n_0\Delta^\pm(k)}\begin{bmatrix}\varphi_0'(x) + \left(1 - \frac{m}{n_0}\right)\omega^2(k) \times \\ \times \int dy \varphi_0'(y) G^\pm(k, x-y)\end{bmatrix}\right\},$$

$$\Delta^{(+)}(k) = 1 + \frac{1}{n_0}\left(1 - \frac{m}{n_0}\right)\omega^2(k)\int dxdy\,\varphi_0'(y)G^{\pm}(k, x-y)$$

$$\Delta^{(-)}(k) = \Delta^{(+)*}(k), \quad G^{\pm}(k, x-y) = \pm\frac{i}{2k}\exp(\pm ik|x-y|).$$

$$\varphi_0(k) = \int dx\,K(x)\varphi_0(x).$$

Taking into account (21), for $p_\lambda$ and $\pi_t(x,t)$ we have:
$$p_\lambda = p_k e^{-i\omega(k)t}, \quad \pi_t(x,t) = \pi_k(x)e^{-i\omega(k)t}.$$

The normalizing factor N is found from the normalization conditions imposed on $\lambda_k, p_{\lambda k}, \varphi_k(x)$ and $\pi_{tk}(x)$:

$$p_{\lambda k}^* \lambda_{k'} + \lambda_k^* p_{\lambda k'} + \int dx\,(\pi_{tk}^*(x)\varphi_{k'}(x) + \varphi_{k'}^*(x)\pi_{tk}(x)) = \delta(k-k')$$

and equals
$$N = [(2\pi)(2\omega(k))]^{-1/2} \equiv N(k).$$

These solutions have the following asymptotic behavior:
$$\varphi_k^{(+)} = N(k)S_+(k)e^{ikx}, \quad \text{as } x \to +\infty, \tag{24}$$
$$\varphi_k^{(+)} = N(k)\left(e^{ikx} + R_+(k)e^{-ikx}\right), \quad \text{as } x \to -\infty, \tag{25}$$
$$\varphi_k^{(-)*} = N(k)\left(e^{-ikx} + R_-(k)e^{ikx}\right), \quad \text{as } x \to +\infty, \tag{26}$$
$$\varphi_k^{(-)} = N(k)S_-(k)e^{-ikx}, \quad \text{as } x \to -\infty, \tag{27}$$

where $S_\pm(k)$ and $R_\pm(k)$ are the reflection and transmission coefficients correspondingly of the form:

$$S_+(k) = 1 - \frac{ik\varphi_0^2(k)}{2n_0\Delta^+(k)}\left(1 - \frac{m}{n_0}\right)\omega^2(k) \tag{28}$$

$$R_+(k) = \frac{ik\varphi_0^2(k)}{2n_0\Delta^{(+)}(k)}\left(1 - \frac{m}{n_0}\right)\omega^2(k)$$

$$S_-(k) = 1 - \frac{ik\varphi_0^2(k)}{2n_0\Delta^{(-)*}(k)}\left(1 - \frac{m}{n_0}\right)\omega^2(k)$$

$$R_-(k) = \frac{ik\varphi_0^2(k)}{2n_0\Delta^{(-)*}(k)}\left(1 - \frac{m}{n_0}\right)\omega^2(k).$$

Taking into account the equation

$$\frac{ik\varphi_0^2(k)}{2n_0}\left(1-\frac{m}{n_0}\right)\omega^2(k) = \frac{i}{2k}(\Delta^{+*}(k) - \Delta^+(k))$$

one obtains:

$$S_-(k) = S_+(k) = \cos\alpha\, e^{-i\alpha(k)}, \quad R_-(k) = R_+(k) = \sin\alpha\, e^{-i\alpha(k)}, \tag{29}$$

where

$$\Delta^+(k) = |\Delta^+(k)|e^{i\alpha(k)}.$$

### 3. S-matrix

Matrix element of the S-matrix between states $\Psi_i(a(t_i), \lambda(t_i), \varphi(x, t_i))$ and $\Psi_f(a(t_f), \lambda(t_f), \varphi(x, t_f))$ is defined by the expression:

$$S_{fi} = \lim_{\substack{t_f \to \infty \\ t_i \to -\infty}} \int da(t_i)da(t_f)d\lambda(t_i)d\lambda(t_f)d\,\varphi(x,t_i)d\varphi(x,t_f)\Psi_f^*\Psi_i \times$$

$$\times \int DaDp_a D\lambda Dp_\lambda D\varphi(x,t)D\pi_t(x,t)\delta(\psi)\delta(\chi)|\{\psi,\chi\}| \times$$

$$\times \exp\left(i\left[\int_{t_i}^{t_f} dt(p_a\dot{a} + p_\lambda\dot{\lambda}) + \int dx\pi_t(x.t)d\varphi(x,t) - H.\right]\right) \tag{30}$$

By virtue of the momentum conservation law $\Psi(a, \lambda, \varphi(x, t))$ can be represented as

$$\Psi(a, \lambda, \varphi(x,t)) \to \exp(iap_a)\Psi(\lambda, \varphi(x,t))$$

which allows us to explicitly integrate over the variables $a$ and $p_a$ in (30) and to isolate the β-function indicating the exact fulfillment of the conservation law:

$$S_{fi} = 2\pi\delta(p_{a_f} - p_{a_i})\lim_{\substack{t_f \to \infty \\ t_i \to -\infty}}\int d\lambda(t_i)d\lambda(t_f)d\,\varphi(x,t_i)d\varphi(x,t_f)\Psi_f^*\Psi_i \times$$

$$\times \int D\lambda Dp_\lambda D\varphi(x,t)D\pi_t(x,t)\delta(\psi)\delta(\chi)|\{\psi,\chi\}| \times$$

$$\times \exp\left(i\left[\int_{t_i}^{t_f} dt p_\lambda\dot{\lambda} + \int dx\pi_t(x.t)d\varphi(x,t) - H.\right]\right)$$

To calculate (31), we write the generating functional S of the matrix in a holomorphic representation in the center-of-mass system. Therefore, we introduce the creation and annihilation operators [12]:

$$\lambda = \frac{1}{\sqrt{2m\omega}}(a^+(t) + a(t)), \quad p_\lambda = i\sqrt{\frac{m\omega}{2}}(a^+(t) - a(t)), \tag{32}$$

$$\varphi(x,t) = \int dxk\left(\varphi_{0k}^*(x)b^+(k,t) + \varphi_{0k}(x)b(k,t)\right),$$

$$\pi_t(x,t) = i\int dk(\pi_{t0k}^*(x)b^+(k,t) - \varphi\pi_{t0k}(x)b(k,t)),$$

which lead (19) to the following expression

$$H = \omega a^+(t)a(t) + \int dk\omega(k)b^+(k,t)b(k,t)\left(a^+(t)+a(t)\right) \times$$
$$\times \int dk(\eta(\kappa)b^+(k,t) + \eta_\kappa^* b(k,t)) \qquad (33)$$

in which $\eta_\kappa = -\frac{1}{\sqrt{2m\omega}}\int dx K' \varphi_{0k}^*(x)$ with $\varphi_{0k}(x)$ to be a solution of the equation

$$(-\partial_x^2 - k^2)\varphi_{0k}^{(\pm)}(x) = \frac{\varphi_0'(x)}{n}\int dy\, \varphi_0'(y)(-\partial_y^2 + \mu^2)\varphi_{0k}^{(\pm)}(y) \qquad (34)$$

and is

$$\varphi_{0k}^{(\pm)}(x) = N_0(k)\left[e^{ikx} + \frac{ik\varphi_0(k)}{\Delta_0^\pm(\kappa)}\int dy G^{(\pm)}(k,x-y)\,\varphi_0'(y)\right] \qquad (35)$$

with $\Delta_0^\pm(\kappa) = \int dxdy\varphi_0'(x)G^{(\pm)}(k,x-y)\,\varphi_0'(y)$. The quantity $\pi_{t0k}(x) = \omega(k)\varphi_{0k}(x)$. These functions are normalized by the condition:

$$\int dx\left(\pi_{t0k}^*(x)\varphi_{0k'}(x) + \varphi_{0k}^*(x)\pi_{t0k'}(x)\right) = \delta(\kappa - \kappa')$$

which makes it possible to find the normalization factor $N_0(k) = [(2\pi)(2\omega(k)]^{-1/2}$. Note that $a(t), a^+(t), b(k,t)$ and $b^+(k,t))$ satisfy the following Dirac brackets: $\{a(t), a^+(t)\}_D = -i$, $\{b^+(k,t), b(k,t)\}_D = -i\delta(k-k')$. The remaining brackets are equal to zero. Based on the results of [13], it can be shown that the generating functional in the zeroth approximation takes the form:

$$\langle a^+(t_f) b^+(k, t_f) | S_0 | a(t_i) b(k, t_i) \rangle = \tag{36}$$

$$= \lim_{\substack{t_f \to \infty \\ t_i \to -\infty}} \frac{1}{2\pi} [G(t_f, a^+(t_f) | 0, 0) G(0, 0 | t_i, a(t_i))]^{-1} \times$$

$$\times \int \exp\left\{\frac{1}{2} [a^+(t_f) a(t_f) + a^+(t_i) a(t_i)] + \right.$$

$$+ \frac{1}{2} \int dk [b^+(k, t_f) b(k, t_f) + b^+(k, t_i) b(k, t_i)] \left.\right\} \times$$

$$\times \exp i \int_{t_i}^{t_f} dt \left\{ \frac{1}{2i} (\dot{a}^+(t) a(t) - a^+(t) \dot{a}(t)) - \omega a^+(t) a(t) + \right.$$

$$+ \int dk \left[ \frac{1}{2i} (\dot{b}^+(k, t) b(k, t) - b^+(k, t) \dot{b}(k, t)) - \omega(k) b^+(k, t) b(k, t) - \right.$$

$$- (a^+(t) + a(t))(\eta_k b^+(k, t) + \eta_k^* b(k, t)) \left.\right] \right\} \times$$

$$\times Da(t) Da^+(t) Db(k, t) Db^+(k, t),$$

in which

$$a(t_i) = a e^{-i\omega t_i}, \quad a^+(t_f) = a^+ e^{i\omega t_f}, \tag{37}$$

$$b(k, t_i) = b(k) e^{-i\omega t_i}, \quad b^+(k, t_f) = b(k)^+ e^{i\omega t_f}, \tag{38}$$

Второй интеграл выражения (36) является гауссовым, поэтому он легко вычисляется и равен подынтегральному выражению, вычисленному в точке эктремума показателя экспоненты. Условия экстремумов выглядят следующим образом:

$$\dot{b}^+(k, t) - i\omega(k) b^+(k, t) - i\eta_\kappa^* A(t) = 0 \tag{39}$$

$$\dot{b}(k, t) + i\omega(k) b(k, t) + i\eta_k A(t) = 0 \tag{40}$$

$$\dot{a}^+(t) - i\omega a^+(t) - iB(t) = 0 \tag{41}$$

$$\dot{a}(t) + i\omega a(\dot{t}) + iB(t) = 0 \tag{42}$$

in which he following notations are introduced:

$$A(t) = a^+(t) + a(t) = A^*(t) \tag{43}$$

$$B(t) = \int dk (\eta_\kappa b^+(k, t) + \eta_k^* b(k, t)) = B^*(t). \tag{44}$$

The formal solutions of equations (39) – (42) taking into account equations (37) and (38) are as follows:

$$b^+(k, t) = b^+(k) e^{i\omega(k) t} - i\eta_k^* e^{i\omega(k) t} \int d\tau \, e^{-i\omega(k)\tau} A(\tau), \tag{45}$$

$$b(k,t) = b(k)e^{-i\omega(k)t} - i\eta_k e^{-i\omega(k)t} \int d\tau\, e^{i\omega(k)\tau} A(\tau), \quad (46)$$
$$a^+(t) = a^+ e^{-i\omega t} - ie^{-i\omega t} \int d\tau\, e^{+i\omega\tau} B(\tau), \quad (47)$$
$$a(t) = ae^{+i\omega t} - ie^{+i\omega t} \int d\tau\, e^{-i\omega\tau} B(\tau) \quad (48)$$

Due to the quadratic nature of the integrand of the second factor (36), all terms, except the boundary ones, do not contribute to the above integral. Substituting (47) and (48) into $\frac{1}{2}(a^+(t_f)a(t_f) + a^+(t_i)a(t_i))$ and into (43), we obtain:

$$\frac{1}{2}\big(a^+(t_f)a(t_f) + a^+(t_i)a(t_i)\big) =$$
$$= a^+ a - \frac{i}{2}(a^+ + a)\int_{t_i}^{t_f} d\tau \cos\omega\tau\, B(\tau) - \frac{i}{2}(a^+ - a)\int_{t_i}^{t_f} d\tau \sin\omega\tau\, B(\tau),$$

$$\mathrm{Im}A(t) = (a^+ - a)\sin\omega t - \int_{t_i}^{t_f} d\tau \cos\omega(t-\tau) B(\tau) = 0,$$

$$\frac{d}{dt}\mathrm{Im}A(t) = (a^+ - a)\cos\omega t + \int_{t_i}^{t_f} d\tau \sin\omega(t-\tau) B(\tau) = 0$$

At time t=0 we will have
$$\int_{t_i}^{t_f} d\tau \cos\omega\tau\, B(\tau) = 0, \quad (a^+ - a)\sin\omega t = 0. \quad (49)$$

Thus, for the second factor (36) we have:
$$\exp\left\{a^+ a - \frac{i}{2}(a^+ - a)^2\right\} \exp\left(\left\{\frac{1}{2}\int dk[b^+(k,t_f)b(k,t_f) + b^+(k,t_i)b(k,t_i)]\right\}\right) \quad (50)$$

Here
$$b(k,t_f) = [b(k)S_+(k)\theta(k) + b(k)S_-(k)\theta(-k) +$$
$$b(-k)R_-(k)\theta(k) + b(-k)R_+(-k)\theta(-k)]e^{-i\omega(k)t_f},$$
$$b^+(k,t_i) = [b^+(k)S_+(k)\theta(k) + b^+(k)S_-(k)\theta(k) +$$
$$b^+(-k)R_+(k)\theta(k) + b^+(-k)R_-(-k)\theta(-k)]e^{-i\omega(k)t_i},$$

For the exponent of the second factor in (50) we obtain:

$$\int dk[\, b(k)b^+(k)S_+(|k|) + b(-k)b^+(k)R_+(-k)]. \quad (51)$$

For the matrix element of the generating integral (36), taking into account (49), (51) and (29), we finally obtain:

$$S_{0,fi} = e^{-i\alpha(k_1)}[\delta(k_1 - k_2)\cos\alpha(k_1) + i\,\delta(k_1 + k_2)\sin\alpha(k_1)]. \quad (52)$$

## Conclusion

The Bogolyubov transformations in strong coupling theory led to a singular Lagrangian containing constraints. Using the Dirac formalism and a modified Feynman integral, we were able to construct $S-$ matrix that correctly describes the symmetry properties of the system under consideration. The advantage of this method is that in this formulation of the tight-binding method for constructing the Hamiltonian, there is no need to fix an additional condition from the very beginning, as is done in [10] and other formulations. In addition, we were able to obtain an explicit expression for the total momentum of the system, and the effective mass of the part is equal to $M_0 + n_0$, i.e., due to the interaction, the particle acquired a mass $n_0$.